\documentclass[review,number,sort&compress]{elsarticle}

\usepackage{amssymb}
\usepackage{epsfig}
\usepackage{graphicx}
\usepackage{bm}
\usepackage{dcolumn}
\usepackage{lineno}
\usepackage{physics}
\usepackage{array, multirow}
\usepackage{amsmath}
\usepackage{lineno}

\journal{NIM-A}

\begin{document}

\begin{frontmatter}



\title{Systematic Uncertainties in RF-Based Measurement of Superconducting Cavity Quality Factors}


\author{J.P. Holzbauer\corref{ref1}}
\ead{jeremiah@fnal.gov}
\author{Yu. Pischalnikov}
\author{D.A. Sergatskov}
\author{W. Schappert}
\author{S. Smith}
\cortext[ref1]{Operated by Fermi Research Alliance, LLC under Contract No. De-AC02-07CH11359 with the United States Department of Energy.}
\address{Fermi National Accelerator Laboratory\\P.O. Box 500, Batavia, IL 60510, USA}

\begin{abstract}
$Q_0$ determinations based on RF power measurements are subject to at least three potentially large systematic effects that have not been previously appreciated. Instrumental factors that can systematically bias RF based measurements of $Q_0$ are quantified and steps that can be taken to improve the determination of $Q_0$ are discussed. 
\end{abstract}


\end{frontmatter}


\section{Introduction}

The intrinsic quality factor, $Q_0$, of a superconducting cavity is an important measure of its performance. The ability to produce cavities with higher $Q_0$ could reduce capital and operating costs of future accelerators.  Research into both the fundamental superconducting properties and preparation techniques required to achieving high quality factors is ongoing at many institutions~\cite{SRF_res1, SRF_res2, SRF_res3}. To fully understand how cavity performance might be improved, systematic uncertainties in the measurements used to extract material properties must be well understood~\cite{CornellSys}.

If the intrinsic coupling factor between cavity and coupler, $\beta$, is close to unity during testing $Q_0$ can be determined from direct measurement of RF losses in the cavity \cite{Powers}. On the other hand, if the coupling is far greater than unity, cryogenic heat load measurements must be employed. Only RF measurement techniques will be considered here.

\begin{figure}[!ht]
   \centering\includegraphics[clip=true,trim=1cm 4cm 0cm 2cm,width=130mm]{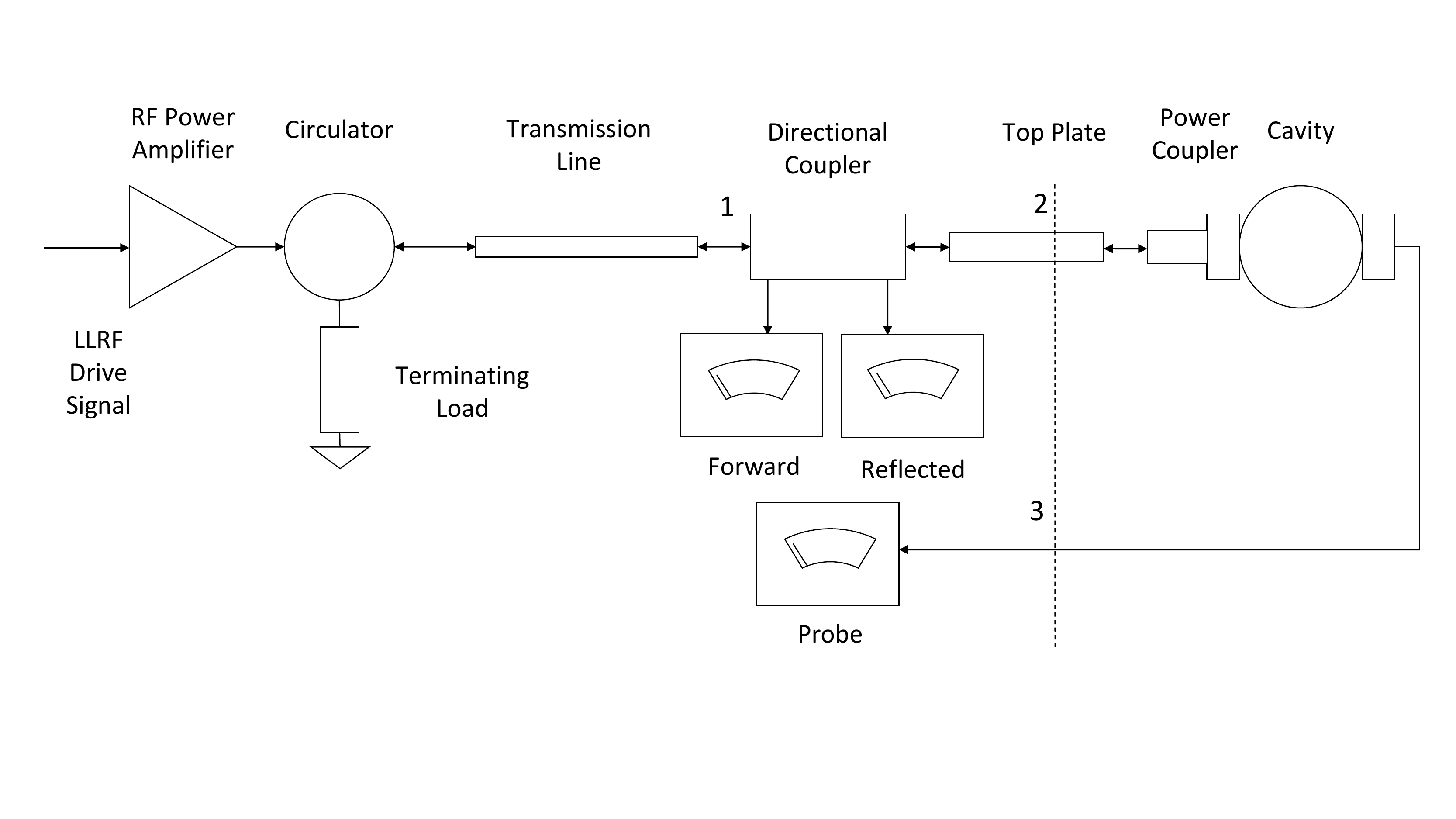}
   \caption{Simplified schematic of the Fermilab Vertical Test Stand RF measurement system.}
   \label{RFSchematic}
\end{figure}

RF-based quality factor measurements commonly employ a circuit similar to that shown schematically in Figure~\ref{RFSchematic}. The cavity is excited by a CW drive signal via a power antenna with coupling close to matched and the cavity field is monitored by a weakly coupled probe antenna. The forward, reflected, and transmitted powers: $P_{F}$, $P_{R}$, $P_{T}$ are measured during steady state operation and the cavity decay time, $\tau$, is measured when the power to the cavity is shut off.

The loaded quality factor, $Q_L$, can be determined from the angular frequency of the RF drive waveform, $\omega$, and from the characteristic decay time, $\tau$, of the stored energy when power to the cavity is shut off:

\begin{equation}
Q_L=\omega \tau.
\end{equation}

Cavity quality factors (and hence the cavity decay time) in general depend on gradient. The decay time can be defined more precisely as the tangent of the decay curve at the beginning of the decay:

\begin{equation}
\tau = -\left.\left(\frac{d \ln(U(t))}{dt}\right)^{-1}\right|_{t_{Decay} = 0}.
\end{equation}

A common practice is to capture and fit the first 10\% of the decay to calculate $\tau$. 

The cavity coupling can be determined by comparing the power dissipated in the cavity, $P_{D}$, to the reactive power, $P_{X}=\omega U = Q_L P_{F}$, when the cavity is precisely on resonance:
\begin{equation}
\beta = \frac{P_{X}}{P_D} = \frac{P_{X}}{P_F - P_R - P_T}.
\end{equation}

The intrinsic quality factor can be determined from the cavity coupling and loaded quality factor as follows:
\begin{equation}
Q_0=(1+\beta) Q_L
\end{equation}
\begin{equation}
Q_{Ext}=\left(1+\beta^{-1}\right)Q_L.
\end{equation}

In practice a measured coupling, $\beta^*$, is determined by comparing the ratio on resonance of the reflected to forward power measurements:

\begin{equation}
\beta^* = \frac{P_{X}}{P_F-P_R}=\left(\frac{1+\sqrt{\frac{P_R}{P_F}}}{1-\sqrt{\frac{P_R}{P_F}}}\right)^{\pm 1}.
\end{equation}

The sign of the exponent in this equation is chosen to be positive (negative) if the cavity is over-coupled (under-coupled).

The coupling of the probe antenna is typically chosen to be much smaller $(<0.1)$ than the coupling of the power antenna. In this case $P_F - P_R \ll P_T$ and the difference between $\beta$ and $\beta^*$ is small:

\begin{equation}
\beta \approx \beta^* \left(1+\frac{P_T}{P_F-P_R}\right)
\end{equation}

RF power levels and cavity decay times can typically be determined with an accuracy of a few percent. 
If uncertainties in $\beta$ and $\tau$ are independent the resulting uncertainty in $Q_0$ can be estimated using standard statistical methods for the propagation of uncertainties~\cite{errorprop}:

\begin{equation}
\sigma_{Q_0}^2 = \left|\frac{\partial Q_0}{\partial \beta}\right|^2\sigma_{\beta}^2+\left|\frac{\partial Q_0}{\partial \tau}\right|^2\sigma_{\tau}^2.
\end{equation}

This leads to an uncertainty in $Q_0$ of: 

\begin{equation}
\left\langle\left(\frac{\Delta Q_0}{Q_0}\right)^2 \right\rangle^{\frac{1}{2}}=\left(\left\langle\left(\frac{\Delta \tau}{\tau}\right)^2\right\rangle+\frac{1}{\left(1+\beta^{-1} \right)^2} \left\langle\left(\frac{\Delta \beta}{\beta}\right)^2 \right\rangle\right)^{1/2}.
\end{equation}

Even under ideal conditions, quality factor measurements using this approach to are limited to accuracies of 5\% or more \cite{Powers, Melnychuk}. 

Implicit in this approach, however, are three assumptions:
\begin{enumerate}
\item The forward and reflected waveforms are perfectly separated by the directional coupler during the coupling factor measurement.
\item No power is incident on the cavity during the decay time measurement.
\item The cavity is precisely on resonance during the coupling factor measurement.
\end{enumerate}

Each of these three assumptions is violated in practice. 

\begin{enumerate}
\item The imperfect directivity of the directional coupler used to separate the waveform incident on the cavity from the reflected waveform inevitably introduces some degree of cross-contamination between the signals.
\item Energy emitted into the reflected waveform from the cavity during the decay can re-reflect back from the circulator commonly used to isolate the RF power amplifier as energy incident on the cavity. The re-reflected energy may interfere constructively or destructively with the cavity field. This interference will systematically bias measured decay times.
\item Energy re-reflected from the circulator will also systematically shift the resonance frequency of the cavity-waveguide system from the true resonance of the cavity leading to systematic biases in the measured coupling factor.
\end{enumerate}

Each of these three effects introduces additional uncertainties in $Q_0$ measurements that may be comparable to or larger than uncertainties associated with power meter calibration and decay time measurements. In the following, direct measurements, analytic calculations, and numerical simulations will be used to quantify uncertainties introduced by each of these effects. Steps that can be taken to reduce uncertainties from these sources will also be outlined. 

\section{Power Meter Calibration and Decay Time Measurement Uncertainties}

Systematic uncertainties in $Q_0$ measurements from the calibration of the power meters used to monitor the cavity signals and from cavity decay time measurements have been discussed in detail elsewhere \cite{Powers, Melnychuk} but will be briefly outlined here for completeness.

If the fractional uncertainties in the calibration of each power meter (forward, reflected, and probe) are assumed to be the same, the uncertainty in the measured coupling factor is given by the following expression:

\begin{equation} 
\left\langle\left(\frac{\Delta \beta^*}{\beta^*}\right)^2\right\rangle^{\frac{1}{2}}_{PM} = \frac{\sqrt{2}}{4}\left|\beta^*-{\beta^*}^{-1}\right|\left\langle\left(\frac{\Delta P}{P}\right)^2\right\rangle^{\frac{1}{2}}
\end{equation}

\begin{figure}[!ht]
   \centering\includegraphics[clip=true,trim=1cm 6cm 2cm 7cm,width=120mm]{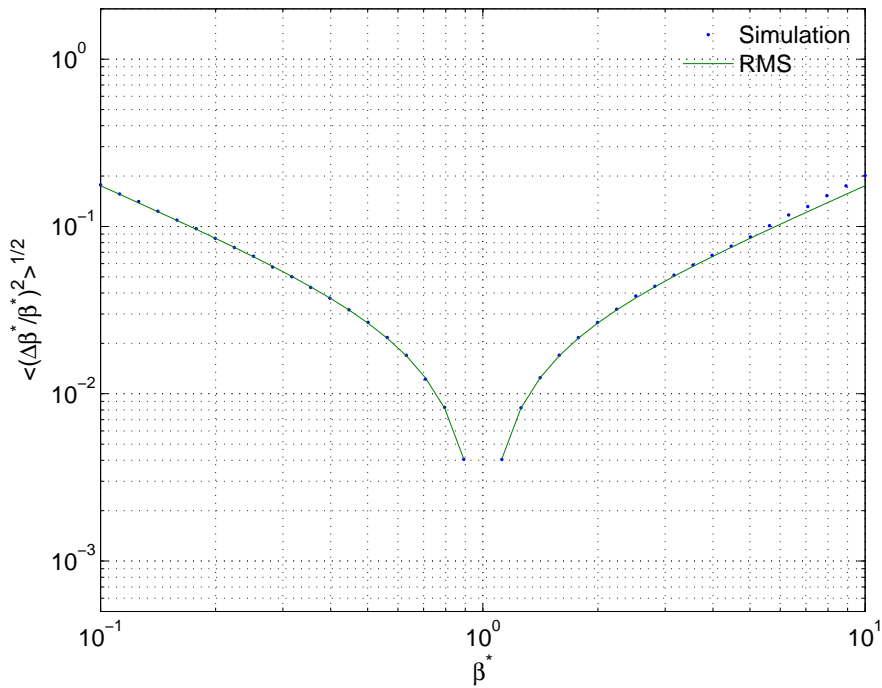}
   \caption{Systematic uncertainties associated due to power meter calibration.}
   \label{PMcalErrors}
\end{figure}

Figure~\ref{PMcalErrors} shows the systematic uncertainty in the measured coupling factor as a function of coupling factor. The first-order analytic expression for the RMS uncertainty (green line) agrees well with Monte Carlo simulations (blue dots) over most of the range.  The red line shows the peak uncertainty. As $\beta^*$ becomes larger ($\beta^*\rightarrow10$) the simulation results exceed the analytic estimates, indicating the analytic expression under-estimates the uncertainty for large values of $\beta$.

A previous analysis has estimated decay time measurement can be measured to an accuracy of 3\% \cite{Melnychuk}. Additional systematic effects associated with energy reflected back into the forward wave by circulator impedance mismatches were not considered in that analysis will be discussed in detail below.

\section{Directivity Uncertainties}

Dual directional couplers are commonly used to separate the voltage incident on the cavity from the voltage in the waveform reflected from the cavity. Perfect separation of the forward and reflected waves within the coupler is not possible. The level of cross-contamination that may be expected is specified by the directivity of the coupler. The directivity of the forward port can be determined from the S-parameters of the coupler. 

\begin{equation}
D = 20\log_{10}\frac{S_{31}}{S_{41}}
\end{equation}

Poor directivity couplers may have directivities as low as 10 dB. Couplers with directivities of 20 or 30 dB are commonly employed for cavity testing. Ultra-high directivity couplers may have values as high as 60 dB.

While a directivity of 20 dB implies that less than one percent of the power is leaking into the other port, depending on the relative phases of the direct signal and the contamination, interference effects can lead to systematic power mismeasurements of up to $\pm$10 percent. The measured power may be systematically deviate from than the true power by up to $10^{-D/20}$. 

\begin{figure}[!ht]
   \centering\includegraphics[clip=true,trim=0cm 3cm 0cm 0cm,width=130mm]{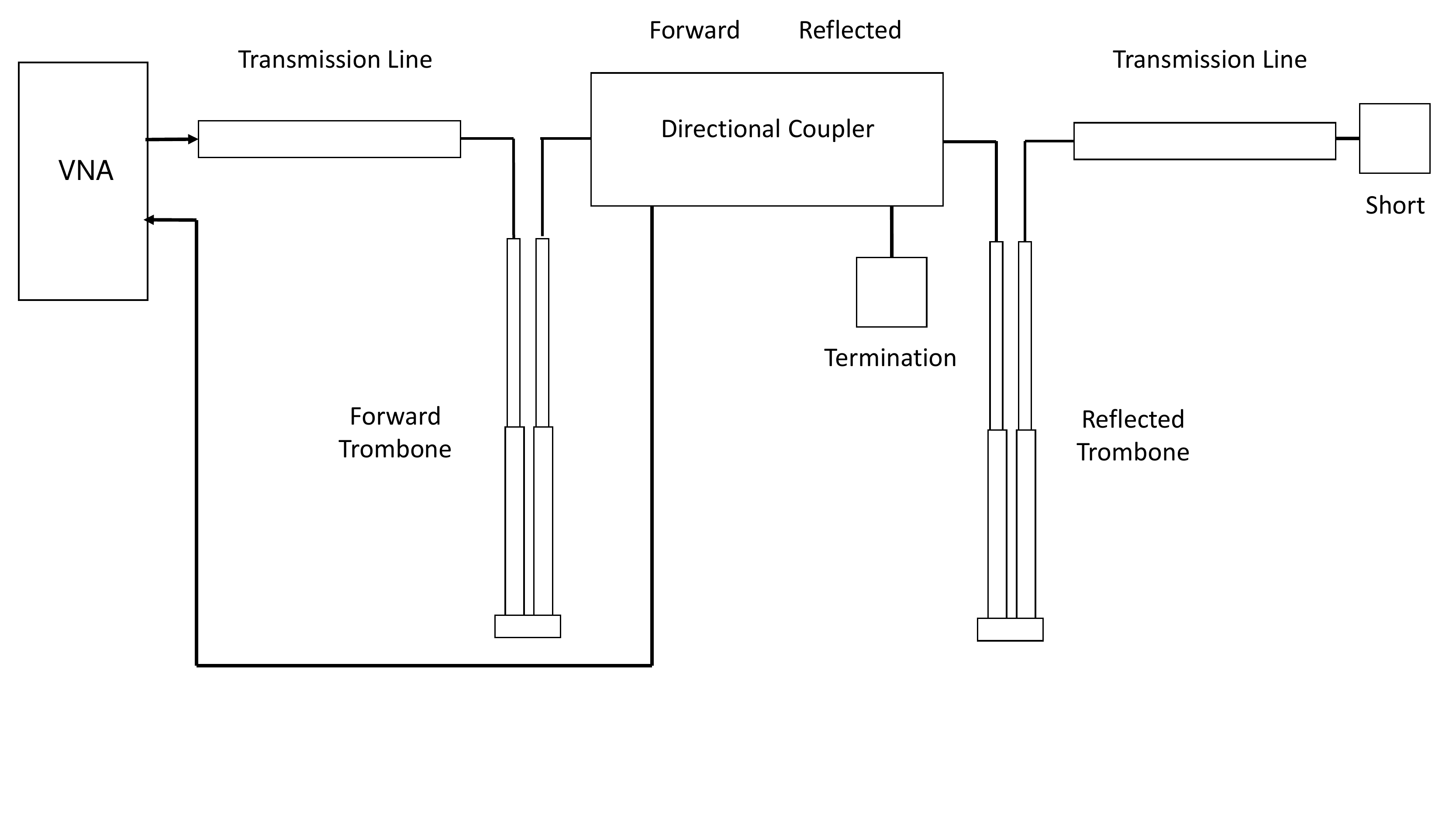}
   \caption{Directivity measurement system.}
   \label{DirSetup}
\end{figure}

To demonstrate this, a variable-length rigid coaxial airline (trombone) was inserted on either side of a directional coupler. The circuit (seen in Figure~\ref{DirSetup}) was driven by a Vector Network Analyser (VNA) and terminated with either a short or an open. The relative transmission from the VNA to the forward power port on the directional coupler was measured as the trombones were swept to independently change both the length of the cables between the VNA and the coupler and between the coupler and the termination. 

\begin{figure}[!ht]
   \centering\includegraphics[clip=true,trim=1cm 6cm 2cm 7cm,width=130mm]{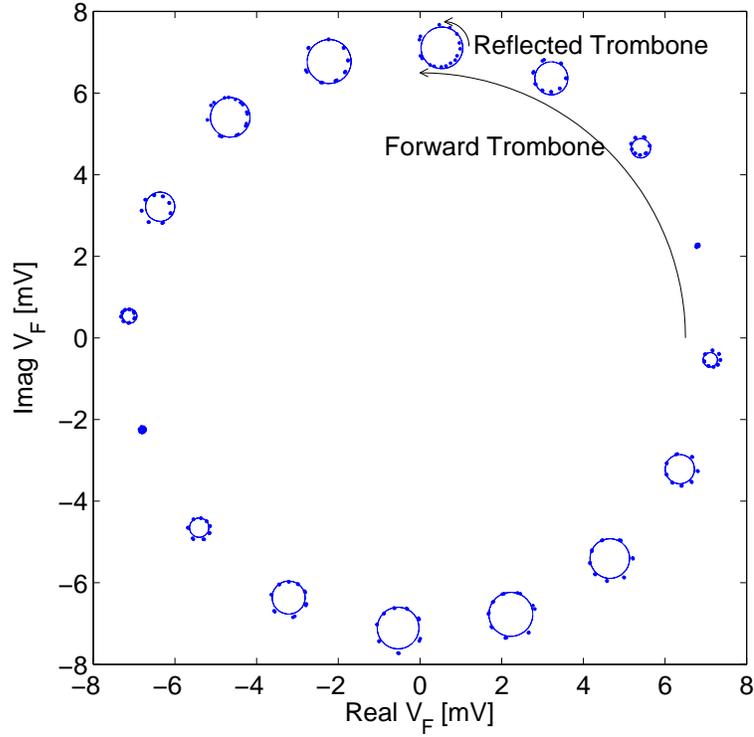}
   \caption{Direct directivity measurement including reflections from the VNA. Dots are measured values, circles are the fit as described. Measurement of an HP776D Duel Directional Coupler.}
   \label{DirMeasure}
\end{figure}

The complex signals measured by the VNA from the forward port of the coupler measured as the lengths of the two trombones was varied are shown as dots in Figure \ref{DirMeasure}. As the length of the trombone between the VNA and the coupler changed, the phase of the direct component of the forward signal swept through a large circle in the complex plane. As the length of the trombone between the coupler and the termination changed, the cross-contamination of the forward from the reflected signal sweeps around a smaller circle centered at the value of the direct forward signal.  The radii of the smaller circles vary because some fraction of the signal reflected from the termination is reflected back into the forward direction by the VNA.  That reflected wave can interfere positively or negatively with the contamination. The measurement was fit assuming directivity contamination $\epsilon_F$, reflections from the termination (short or open) $\Gamma_{S/O}$, and reflections from the VNA $\Gamma_{VNA}$ with the form:

\begin{equation}
V_F^{Measured} = V_F e^{i\phi_F}+\epsilon_F\Gamma_{S/O}V_F e^{i(\phi_F+2\phi_R)}+\Gamma_{S/O}\Gamma_{NWA}V_Fe^{i(3\phi_F+2\phi_R)}
\end{equation}

The results of the fit are shown as solid lines in Figure \ref{DirMeasure}. The fit reproduced the measured values of the forward signal as the trombone lengths changed to 1.5\%. In this measurement, $\Gamma_{NWA}$ was measured as $-0.038+0.004i$, a reflection coefficient of -28 dB. The contamination constant $\epsilon_F$ was measured to be $0.009+0.036i$, or about a 29 dB directivity, close to the manufacturer's specification of 30 dB. 

Mismeasurements of the forward and reflected power due to imperfect directivity lead to a systematic biases in the cavity coupling factor determined from those measurements.  If directivity is modelled as a linear mixing of forward and reflected signals (voltages), the leading order expressions for forward and reflected powers are:

\begin{eqnarray}
\frac{\Delta P_F}{P_F} \approx 2\frac{1-{\beta^*}^{-1}}{1+{\beta^*}^{-1}}|\epsilon_F|\cos{\theta_F}\\
\frac{\Delta P_R}{P_R} \approx 2\frac{1+{\beta^*}^{-1}}{1-{\beta^*}^{-1}}|\epsilon_R|\cos{\theta_R}
\end{eqnarray}

The magnitudes of the cross-contamination are bounded by the directivity of the coupler:

\begin{eqnarray}
|\epsilon_F| \le 10^{-D/20}\\
|\epsilon_R| \le 10^{-D/20}
\end{eqnarray}

The phase angles, $\theta_F$ and $\theta_R$, of the cross-contamination depend on the construction of the coupler and the length of the transmission line connecting the coupler to the cavity. Unless these phases have been explicitly measured they must be treated as random sources of systematic uncertainty.

\begin{figure}[!ht]
   \centering\includegraphics[clip=true,trim=1cm 6cm 2cm 7cm,width=130mm]{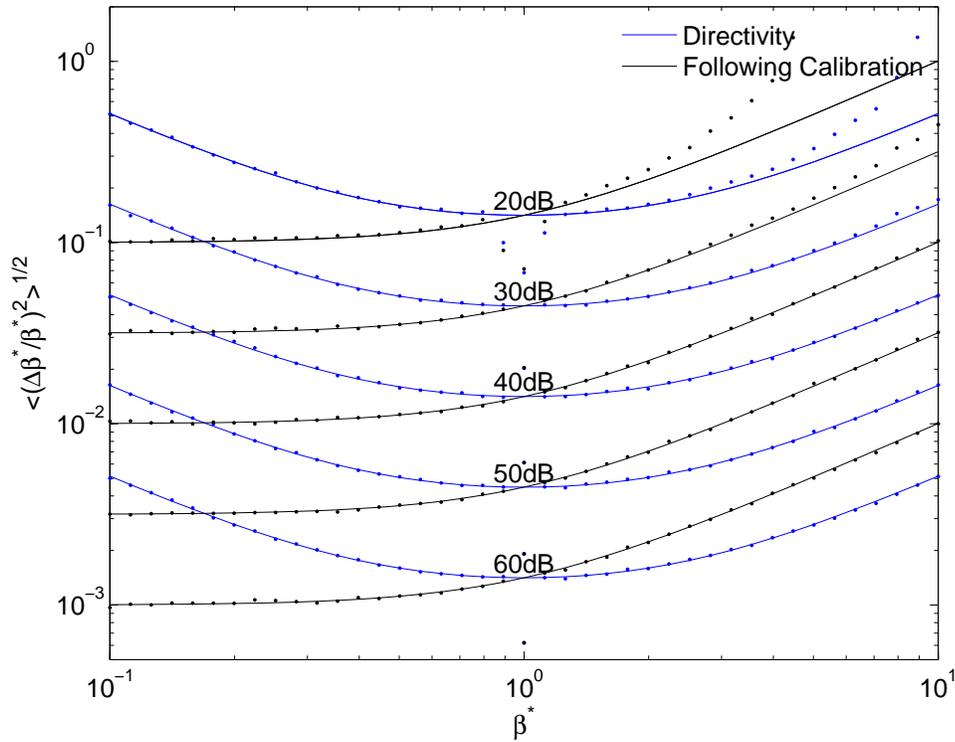}
   \caption{Systematic uncertainties associated with coupler directivity. Dots are MC simulation, lines are linear calculations. Data in blue is for measurement error, and the data in black is this error after the final step of the calibration procedure.}
   \label{CalibratedErrors}
\end{figure}

This will lead to an RMS uncertainty in the measured coupling factor of:

\begin{equation}
\left\langle\left(\frac{\Delta \beta^*}{\beta^*}\right)^2\right\rangle^{\frac{1}{2}}_{PM} \approx 10^{-\frac{D}{20}}\sqrt{1+\frac{\left(\beta^*+{\beta^*}^{-1}\right)^2}{4}}.
\end{equation}

RMS uncertainty as a function of coupling factor are plotted in Figure \ref{CalibratedErrors} for directivities ranging between 20 and 60. For high directivities and coupling factors close to unity, the analytic expression and the simulation results agree well. For poor directivities and high coupling factors the first-order approximation analytic expressions begin to break down and the simulations give higher estimates. For directivities of 20 dB and coupling factors of 10, the fractional uncertainty in $\beta$ can approach or even exceed unity. The black curves in Figure \ref{CalibratedErrors} will be discussed below.

\section{Reflections from the Circulator}

Amplifiers in high power RF circuits are commonly protected by ferromagnetic circulators. Circulators are non-linear devices and rarely present a perfect impedance match to the transmission line connecting the load and the cavity.  Reflections at the mismatch redirect energy from the reverse waveform back into the waveform incident on the cavity. Specifications for ferromagnetic circulators typically quote Voltage Standing Wave Ratios (VSWR) between 1.20 and 1.50. The magnitude of the reflection coefficient and the VSWR are related as follows:

\begin{equation}
\left|\Gamma_{Circulator}\right| = \frac{VSWR-1}{VSWR+1}
\end{equation}

\begin{figure}[!ht]
   \centering\includegraphics[clip=true,trim=1cm 6cm 2cm 7cm,width=110mm]{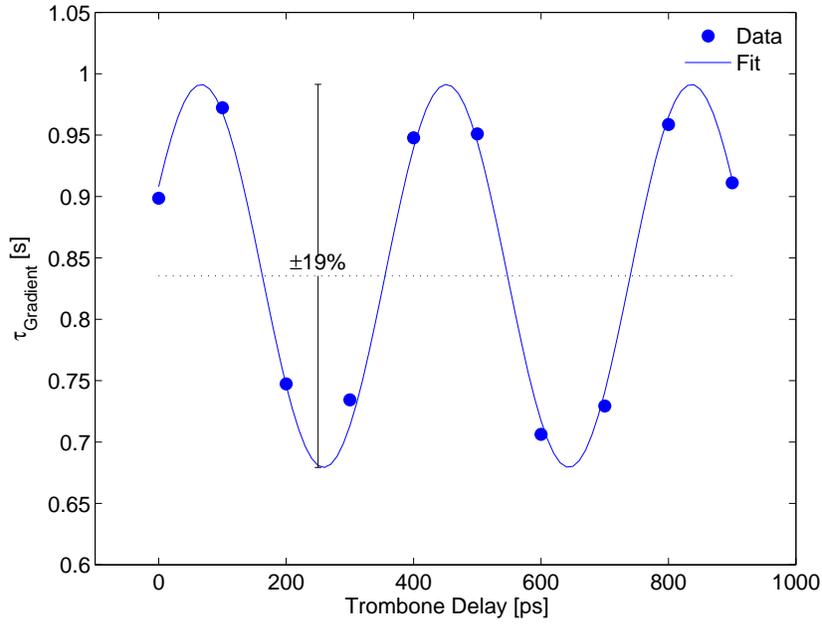}
   \caption{Cavity decay time vs. trombone position as measured on cavity TB9ACC015, a 9-cell 1.3 GHz cavity at FNAL VTS, 7/14/2014.}
   \label{TauVariation}
\end{figure}

These reflections can systematically bias the cavity decay time measurement. Energy re-reflected from the circulator may interfere constructively or destructively with the cavity field. The measured decay time of the cavity will systematically differ from the true cavity decay time depending on the length of the waveguide that connects the cavity and the circulator, $l$, and the wavenumber, $\kappa$, of the RF drive waveform as follows:

\begin{equation}
\frac{\Delta \tau}{\tau} = \frac{2}{1+{\beta^*}^{-1}}\Re\left(\Gamma_{Circulator}e^{-2i\kappa l}\right)
\end{equation}

Constructive interference will systematically bias measured decay times to values longer than the true cavity decay time. Destructive interference will systematically bias the measured decay times to shorter values. 

This can be shown by calculation of the system impulse response. In the frequency domain, the Impulse response takes the form:

\begin{equation}
I_{impulse} = \frac{1}{2\pi}\frac{Z_C}{Z_C+Z_T}\int_{-\infty}^{\infty}d\omega e^{i\omega t}T_{P/F}\frac{e^{-i\kappa l}}{1-e^{-2i\kappa l}\Gamma_{Circulator}\Gamma_{Cavity}}
\end{equation}

where $Z_C$ and $Z_T$ is the circulator and transmission line impedances, $T_{P/F}$ is the cavity transfer function, and $\kappa l$ gives the phase advance between the cavity and circulator. Evaluation with contour integration gives

\begin{equation}
I_{impulse} = \frac{2\omega_{1/2}}{1+\beta^{*-1}}\frac{Z_C}{Z_C+Z_T}\frac{e^{-i\kappa l}}{1+e^{-2i\kappa l}\Gamma_{Circulator}}e^{-\omega^{'}_{1/2}t+i\delta^{'}t}\Theta(t)
\end{equation}

where $\Theta(t)$ represents a Heaviside step function and the prime indicates measured quantities which include circulator reflection effects. For $|\Gamma_{load}|\ll 1$, these measured quantities deviate from the intrinsic values with the following form: 

\begin{equation}
\omega^{'}_{1/2} = \omega_{1/2}+\frac{2\omega_{1/2}}{1+\beta^{*-1}}\Re(\Gamma_{Circulator}e^{-2i\kappa l})
\end{equation}

\begin{equation}
\delta^{'} = \delta-\frac{2\omega_{1/2}}{1+\beta^{*-1}}\Im(\Gamma_{Circulator}e^{-2i\kappa l}).
\end{equation}

From here, the fractional errors in $\tau$ and resonant frequency $\delta$ can be calculated directly.

Figure \ref{TauVariation} shows how the measured decay time changes as the length of a trombone inserted between the cavity and the circulator was varied over one wavelength of the RF drive waveform. As would be expected from the formula given above, the measured decay time oscillates through two full sinusoidal cycles around the true cavity decay time as the length of the trombone sweeps over a wavelength.

\begin{figure}[!ht]
   \centering\includegraphics[clip=true,trim=1cm 6cm 2cm 7cm,width=110mm]{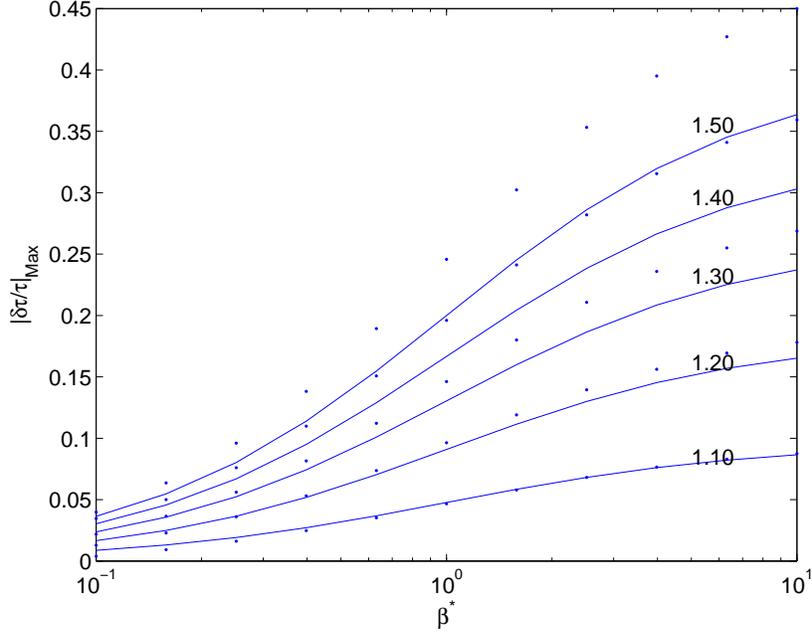}
   \caption{Expected systematic uncertainty the measured cavity decay time with coupling factor and VSWR. Lines are linear calculation, points are full MC calculations.}
   \label{TauvsVSWR}
\end{figure}

If no correction is applied to the measured cavity decay time for energy re-reflected from the circulator the systematic bias in the decay will introduce a systematic bias in $Q_0$.

\begin{equation}
\left\langle\left(\frac{\Delta \tau}{\tau}\right)^2\right\rangle^{\frac{1}{2}}\approx \frac{\sqrt{2}}{1+{\beta^*}^{-1}}\frac{VSWR-1}{VSWR+1}
\end{equation}

A circulator with VSWR of 1.30 will induce probable systematic uncertainty in $Q_0$ of approximately 10\% even for an optimally coupled cavity. Figure \ref{TauvsVSWR} shows how the expected systematic uncertainty in the cavity decay time varies with $\beta^*$ and VSWR. For large values of VSWR and $\beta^*$ the first order equation calculation above (lines) falls below the results of a full simulation (dots).

\section{Resonance Frequency Uncertainties}

Energy re-reflected from the circulator also leads to systematic biases in the measured resonance frequency. The cavity and waveguide together form a coupled resonator system. The peak power transmitted by the cavity/waveguide system is at a frequency systematically offset from the true resonance frequency of the cavity by a factor that depends on the cavity coupling factor, the reflection co-efficient of the circulator, the wavenumber of the drive signal and the length of the waveguide as follows:

\begin{equation}
\frac{\delta_{Cavity} - \delta_{System}}{\omega_{1/2}} = - \frac{2}{1+{\beta^*}^{-1}}\Im\left(\Gamma_{Circulator}e^{-2i\kappa l}\right)
\end{equation}

The formula for the measured cavity coupling factor in Equation 1 assumes that the cavity is perfectly on resonance. If the cavity is mis-tuned because of mismatches at the circulator or imperfect directivity the ratio of reflected to forward power for a detuned cavity will be given by squared magnitude of the ratio of the complex transfer functions:

\begin{equation}
\frac{P_R}{P_F} =\left|\frac{\frac{\beta^*-1}{\beta^*+1}-i\frac{\omega-\delta}{\omega_{1/2}}}{1+i\frac{\omega-\delta}{\omega_{1/2}}}\right|^2
\end{equation}

where $\omega$ is the cavity resonant frequency. The measured coupling factor does not have a simple dependence on detuning so only two limiting cases will be considered here: when the cavity coupling factor is very close unity, and when the coupling factor is far from unity.

When $\beta^*$ is exactly unity and the cavity is exactly on resonance, the reflected power will be exactly zero. As the cavity is detuned from resonance, the reflected power will increase linearly with the magnitude of the detuning. The coupling factor of a detuned cavity will be mis-measured by:
\begin{equation}
\frac{\Delta \beta^*}{\beta^*}\bigg|_{\beta = 1} = \pm 2\left|\frac{\delta}{\omega_{1/2}}\right|.
\end{equation}
The mismeasurement will be positive (negative) if the cavity is over-coupled (under-coupled). 

If the coupling is far from unity, $\left|\beta^*-1\right|>1$, and the cavity is close to resonance, $\left|\delta\omega_{1/2}\right|<1$, the reflected power will depend quadratically on the detuning. In this case the coupling factor will be mis-measured by:
\begin{equation}
\frac{\Delta \beta^*}{\beta^*}\bigg|_{\left|\beta^*-1\right| \gg \left|\delta/\omega_{1/2}\right|} \approx \pm \frac{1}{2}\frac{\beta^*+1}{\beta^*-1}\left(\frac{\delta}{\omega_{1/2}}\right)^2.
\end{equation}
Again, the mismeasurement will be positive (negative) if the cavity is over-coupled (under-coupled). 

\begin{figure}[!ht]
   \centering\includegraphics[clip=true,trim=1cm 6.5cm 2cm 7cm,width=110mm]{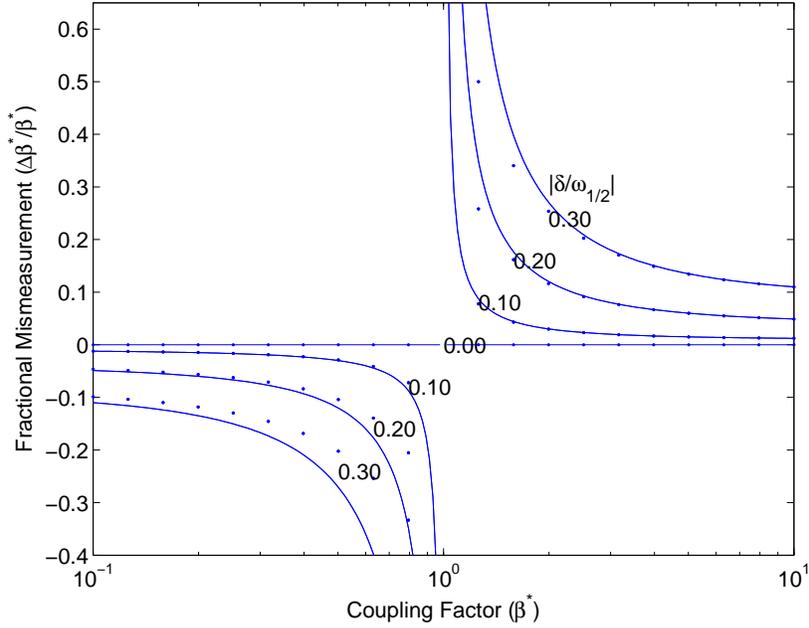}
   \caption{Expected systematic uncertainty the measured cavity resonant frequency with coupling factor and VSWR. Lines are lowest order analytic approximation, points are full numeric calculations.}
   \label{deltavsbeta}
\end{figure}

Figure \ref{deltavsbeta} shows the fractional mis-measurement as a function of coupling factor for a fractional detuning between -0.3 and 0.3. For couplings far from unity, the leading order equation (lines) reproduces the results of a full numerical calculation (points). As expected, for couplings close to unity, the leading order expansion breaks down. In contrast to the other sources of systematic uncertainty discussed here, off-resonance uncertainties are largest when the cavity is close to unity coupling.

\begin{figure}[!ht]
   \centering\includegraphics[clip=true,trim=1cm 6.5cm 2cm 7cm,width=110mm]{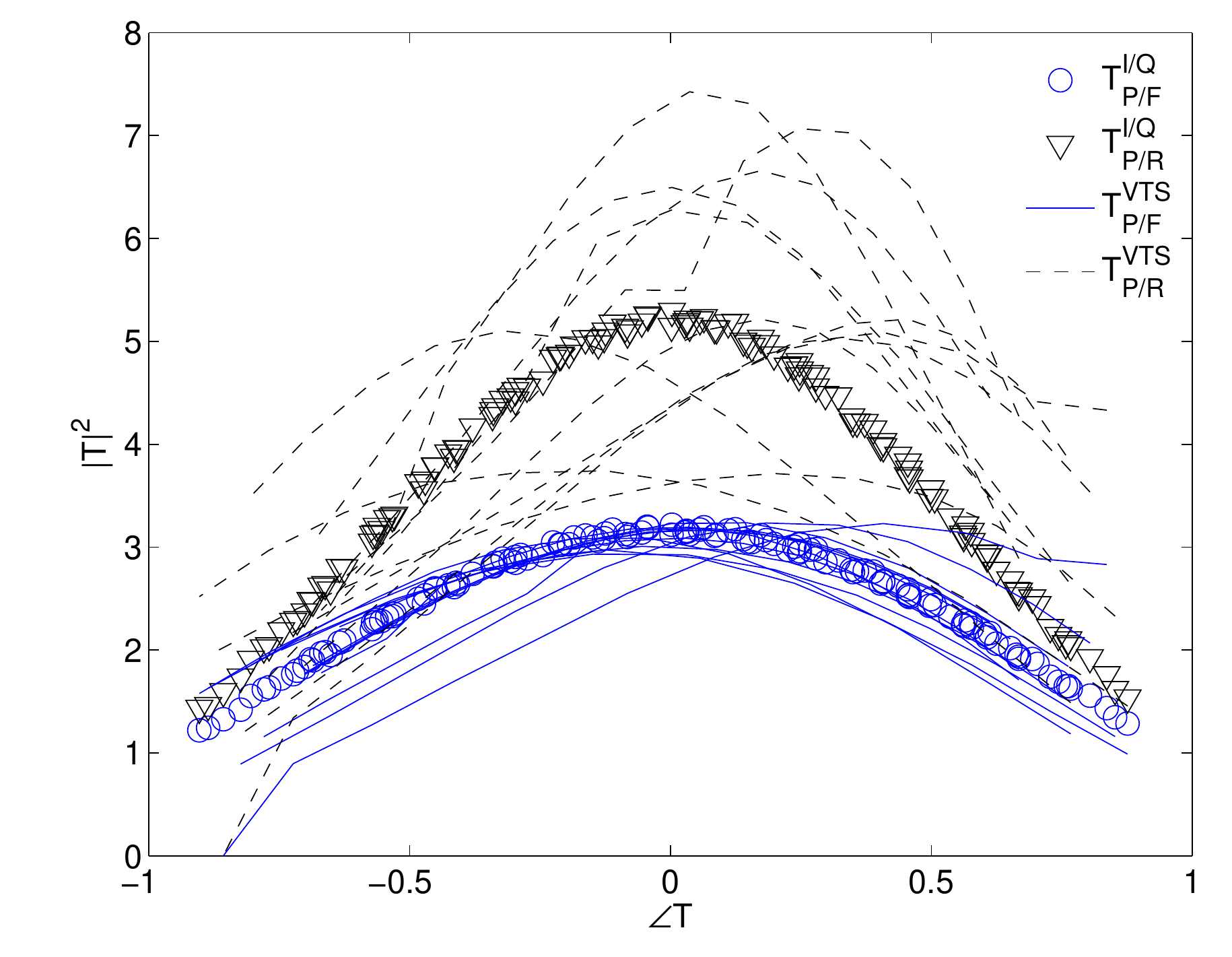}
   \caption{Measured probe/forward (black) and probe/reflected (blue) transfer functions as the length of the waveguide is varied. Triangles/circles are measured with the digital I/Q system, lines/dashes are measured on the analog tracking system. The cavity used, TE1ACC001, was measured on 2/14/2014.}
   \label{DigitalVsAnalog}
\end{figure}

As an illustration, Figure \ref{DigitalVsAnalog} compares the magnitude of the probe/forward and probe/reflected transfer functions measured using Fermilab Vertical Test Stand (VTS) analog tracking system to independent measurements of the same ratios recorded by an independent digital I/Q system. During these measurements, the phase of the analog phase lock loop (PLL) was systematically varied to sweep the RF drive frequency across the cavity resonance. The resonance sweeps were repeated as the length a trombone in the cavity power circuit was varied in steps over a wavelength. The magnitude of each transfer function when plotted against the angle of the transfer function should depend only on the PLL phase and not on the length of the trombone. The transfer functions recorded by the I/Q system (dots) coalesce along a single curve that depends only on the PLL phase and peaks at zero as expected. In contrast the transfer functions recorded by the analog system exhibit significant variations in both magnitude and the peak positions as the length of the trombone is changed. Both directivity and circulator effects could lead to such shifts and no attempt to separate the two was made.

\section{Calibration Uncertainties}

A five step procedure is used to calibrate the power meters in the Fermilab VTS \cite{Powers, Melnychuk}. 

\begin{enumerate}
\item The forward power meter is calibrated to the cryostat top plate by disconnecting the cables from directional coupler and the top plate, injecting a calibrated signal into the directional coupler (Point 1 in Figure \ref{RFSchematic}), measuring the power at the top plate cable (Point 2 in Figure \ref{RFSchematic}) with a portable power meter and comparing it to the forward power meter reading.
\item The reflected power meter is calibrated to the cryostat top plate by disconnecting power cable from the top plate (Point 2 in Figure \ref{RFSchematic}), driving it with a calibrated source, and comparing the reflected power meter to the power of the known source.
\item The probe power meter is calibrated to the cryostat top plate by disconnecting the probe cable from the top plate (Point 3 in Figure \ref{RFSchematic}), driving it with a calibrated source, and comparing the probe power meter to the power of the known source.
\item Losses in the cold cable connecting the cavity field probe to the cryostat top plate are estimated by disconnecting the probe cable from the top plate (Point 3 in Figure \ref{RFSchematic}) with the cavity off resonance, driving the cold cable with a calibrated source through a circulator and comparing the power returned to the circulator load port with the power measured with the second port shorted. 
\item Loses in the cold cable connecting the cavity power coupler to the cryostat top plate are estimated by driving the cavity far off resonance with all cables connected and comparing the readings of the forward and reflected power meters under the assumption that the forward and reflected power are equal and opposite when the cavity is far off resonance.

\end{enumerate}

This calibration procedure is vulnerable to several sources of error. The first four steps of this procedure involve disconnecting and reconnecting cables. Each time the cable configuration is changed the standing wave pattern in the circuit will also change. The magnitude of that change will depend on the quality of the new termination at the end of the cable. Based on measurements in the VTS with a trombone inserted in the cavity field probe cable, the magnitude of the signals may be expected to change by $\pm$ 3\%. This uncertainty may already be reflected in the uncertainties associated with power meter calibration.

The fourth step of the calibration procedure requires disconnecting the cable carrying the field probe signal involves measuring cold cable return loss with a circulator. The circulator may be expected to introduce uncertainties of 10\% to 15\% in the measured attenuation of this cable. This may systematically bias cavity gradient measurements but should have no effect on quality factor measurements since they depend only the forward signal, the reflected signal, and the cavity decay time. 

The final step of the calibration procedure can lead to systematic biases in the $Q_0$. The measured forward and reflected power signals will only be exactly equal and opposite when measured using a perfect directional coupler. The signals measured far off resonance by a coupler with less than perfect directivity will be:

\begin{align}
P_F^{Measured} &= \left(1+2|\epsilon_F|\cos{\theta_F}\right)P_F^{True}\\
P_R^{Measured} &= \left(1+2|\epsilon_R|\cos{\theta_R}\right)P_R^{True}\\
P_F^{Calibrated} &= \left(1-2|\epsilon_F|\cos{\theta_F}\right)P_F^{True}\\
P_R^{Calibrated} &= \left(1-2|\epsilon_R|\cos{\theta_R}\right)P_R^{True}
\end{align}

\begin{equation}
\beta^*_{Calibrated} = \beta^*_{True}\left(1+\frac{|\epsilon_F|\cos\theta_F + |\epsilon_R|\cos\theta_F}{2}\right)
\end{equation}

\begin{equation}
\left\langle\left(\frac{\Delta \beta^*}{\beta^*}\right)^2\right\rangle^{1/2}_{Calibrated}=10^{-\frac{D}{20}}\sqrt{1+\beta^{*2}}
\end{equation}

As can be seen in Figure \ref{CalibratedErrors}, the calibration procedure significantly changes the impact of directivity. Instead of an essentially randomly phased error, calibrating off resonance introduces an additional phase effect that depends on $\beta^*$. In the under-coupled limit ($\beta^*$$\ll$1), moving from far off resonance to resonance doesn't introduce an additional phase shift between forward and reflected. If the cavity is significantly over-coupled ($\beta^*$ $\gg$ 1) however, the phase changes 180 degrees, significantly increasing the systematic errors. Even near matched ($\beta^*$ = 1) this effect is on the same order as the uncalibrated error. Figure \ref{CalibratedErrors} shows both the linear approximation of this error and a simulation of this effect with the full cavity transfer functions. Near matched, this error is suppressed because of the reduction of the magnitude of the reflected power, and in the over-coupled limit, the non-linear terms of the error grow rapidly, especially with poor directivity directional couplers.  

\section{Combined Systematic Uncertainties}

\begin{figure}[!ht]
   \centering\includegraphics[clip=true,trim=1cm 5cm 1cm 6cm,width=110mm]{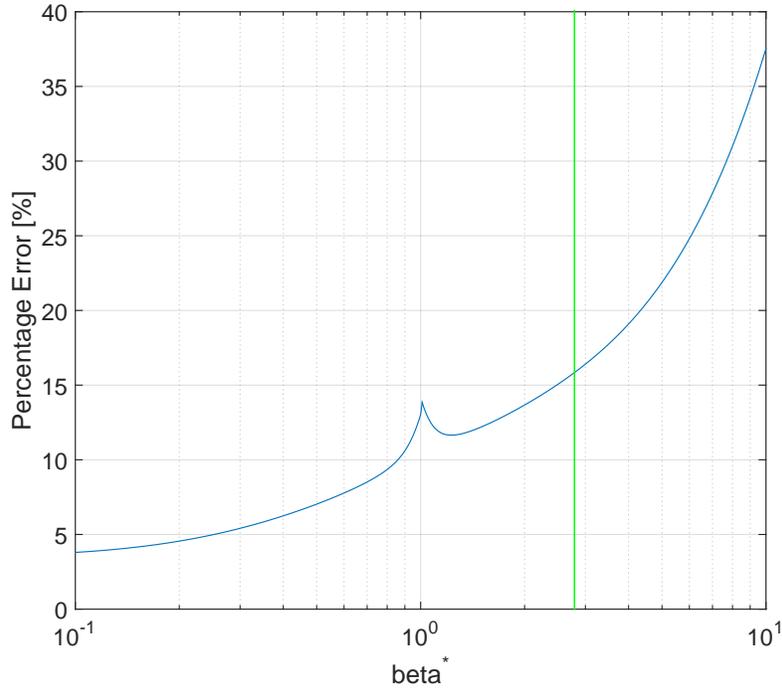}
   \caption{RMS errors in quality factor measurements versus beta, including all sources of error described in this analysis. Assumptions were: VSWR of 1.3, directivity of 30 dB, 3\% error in calculated $\tau$, and 5\% power measurement error. Because of the wide range of $\beta^*$, full numerical calulation of the error contribution from off resonance errors were used. The green line indicates the median beta for all cavities tested at the FNAL VTS between 2007 and 2014.}
   \label{ErrorSum}
\end{figure}

Figure \ref{ErrorSum} shows the combined probable percentage systematic uncertainty in $Q_0$ and its components. The green line on the plot shows the median beta of all single-cell 1.3 GHz cavities tested in the Fermilab VTS between 2007 and 2014.  Three-quarters of those cavities were over-coupled ($\beta^*>$1). The median coupling factor for all cavities was $\beta^*$=2.84.  The probable systematic uncertainty at that coupling factor is just over 15\%. 

\section{Reducing Systematic Uncertainties in Quality Factor Measurements}

Systematic uncertainties in $Q_0$ measurements can be reduced by a variety of measures including:

\begin{enumerate}
\item Using a variable power coupler;
\item Using a high-directivity directional coupler;
\item Using digital I/Q system; 
\item Using data-based calibration; and
\item Measuring complex transfer functions.
\end{enumerate}

Directivity associated uncertainties depend strongly on $\beta^*$ and are smallest when $\beta^*$ is unity. Consistent use of a variable power coupler would allow every measurement to be made while the cavity is optimally coupled and directivity uncertainties are small.

Further improvements can be made by using high-directivity directional couplers. Directional couplers with directivities of 40dB are commercially available. High-directivity couplers may cover narrower frequency bands than broadband couplers or may be limited to lower power levels but if accurate measurements are important, a high-directivity coupler should be employed.

Systematic biases in decay time measurements can be reduced to negligible levels by varying the length of a trombone inserted in the cavity power circuit. Alternatively re-reflected energy can be reduced by installing an impedance matching network at the circulator.

In contrast off resonance errors can only be eliminated if both magnitude and phase data is recorded and the cavity is tuned to the peak of the transfer function rather than the peak probe power or the minimum reflected power as is currently common practice.

Additionally, the existing calibration procedure for the cold drive cable convolves the systematic errors discussed here, increasing measurement error for over-coupled cavities. Direct measurement of the cold cable losses instead of assuming the constraint may reduce error for highly over-coupled cavities. 

\section{Previous Analyses of Systematic Uncertainties}

At least two previous analyses of systematic effects in cavity test stand measurements have been made. 

\begin{enumerate}
\item The designer of the Thomas Jefferson Laboratory VTS upon which the Fermilab VTS is based, presented a tutorial on cavity testing at the 2006 US Particle Accelerator School \cite{Powers}. That tutorial covers both RF and heat load techniques and includes an analysis of systematic effects in RF-based measurements.
\item The analysis in \cite{Powers} was later extended to include correlations between parameters \cite{Melnychuk}. 
\end{enumerate}

Neither of these two analyses considered the phase effects discussed here. As a result, both analyses significantly underestimate the magnitude of the systematic uncertainties in RF-based cavity quality factors measurements.

Impedance mismatches at the circulator were studied earlier at CERN \cite{CERN}, but the tools available at the time did not allow for measurements with the full complex signals.

\section{Conclusion}

$Q_0$ determinations based on RF power measurements are subject to at least three potentially large systematic effects that have not been previously appreciated: directivity, energy re-reflected from the circulator and off-resonance errors. All three of these effects can introduce systematic uncertainties comparable to or larger than the uncertainties associated with power meter calibration and cavity decay times which have been the focus of previous analyses. Measurements of cavity coupling factors can be improved by employing a variable power coupler and installing an impedance matching network to suppress reflections from the circulator. 





\end{document}